# A Toy Model for Cooperative Phenomena in Molecular Biology and the Utilization of Biochemical Applications of PNS in Genetic Applications

S. Bumble*, F. Friedler**, and L. T. Fan***

*Dept. of Physics, Philadelphia Community College, Dept. of Chemistry, Philadelphia, PA 19130, LaSalle University, Philadelphia, PA 19141 & Chestnut Hill College, Philadelphia, PA 19118
**Dept. of Computer Science, University of Veszprém, Hungary
***Dept. of Chemical Engineering, Kansas State University, Manhattan, KS 66506

*"The reigning question in biology is not the survival of the fittest but the arrival of the fittest."*[1]

## Abstract

Qualitative attributes of the region between order and disorder are examined to explore models of genetic and protein networks. Results show how the connectivity of vertices and the strength of their connections are related and how their stability is related to their geometry. It has been possible to relate the interaction energies and chemical potentials of general lattices with the coordination numbers of such lattice models. The results seem to agree with some of those obtained by means of other methods, e.g., those of Barabasi[2,3], Strognatz[4], Dorogovtsev[5], and Kauffman[6] (BSDK). This can yield new perspectives to the cause, treatment and remedies of disease other than the present mode of drug discovery prevalent: the docking of a single ligand on to a target molecule. In order to utilize such results more efficiently, the pathway of biological processes need be elucidated. A method is available for determining biochemical-reaction or metabolic pathways through its systematic synthesis.[7] It is based on a rigorous graph-theoretic method for identifying pathways of catalytic reactions.[8] It synthesizes networks of metabolic pathways using a highly exacting combinatorial method. It generates not only all feasible, independent reaction networks but also those combinations of independent pathways. This method can determine the mechanisms of complex chemical reactions and is applicable to biochemical reactions. It is important to combine this result with the mechanisms existent in gene regulatory networks. Training Genetic Regulatory Networks for feasible biochemical reaction networks or pathways[7] and incorporating such knowledge into DNA would be a superb technique for vanquishing complex diseases.

## Introduction

The mapping of the genetic regulatory network by Kohn[9] has presented a challenge to the theoretician as to the next step. The program "Virtual Cell" from GNS is an application to Kohn's data and his map of the regulatory genetic network at the National Cancer Institute. It can be thought of as an application of the program "Lab View", which is a program from National Instruments that is used to run process control. "A New Kind of Science" by Wolfram[10] presents the idea that any kind of complex problem can be solved by computer programs and he is claiming he will delve more into Molecular Biology. Meanwhile, there has been an explosion of papers on networks and graph theory in the physics archive, arxiv.org on the web. Important works include those of BSDK.

**Biochemical Applications of Process Network Synthesis (PNS)**

A mathematically exact graph-theoretic method proposed by Fan et al[8] for the identification, i.e., determination, of the mechanisms of complex reactions is applicable to biochemical reactions.[7] This method has yielded a complete metabolic pathway network, the maximal structure with minimal complexity, for a given set of candidate elementary reactions. A complete set of feasible sub-networks corresponding to feasible pathways, in turn, can be extracted from the maximal structure. The feasible pathways or mechanisms are to be examined experimentally, computationally and/or theoretically for the final selection of biochemical reactions or mechanism identification. Also, such a method can circumvent the enormous complexities involved in identifying a biochemical or metabolic pathway systematically. This method has been implemented on a PC for the elementary reactions of the conversion of glucose to pyruvate.[7]

The next step can be the examination of genetic networks, genetic regulatory networks, protein networks and signal transduction networks to gain insight into life processes and diseases. In such cases, we can infer that the biomolecules (proteins or assembly of nucleotides) are at the vertices of a network and the "reactions" are the interactions, ligatures or "signals" between the nodes and that the composites lead to the expression of the network(s). It is thought that this approach will lead to results beyond those of the present multitudinous attempts (frequently leading to failure) of researchers at companies and universities today. Such a quasichemical approach has been successful in theoretical physics in the past.

Other methods are presented in the literature; these methods decompose biochemical networks[11], but not necessarily metabolic networks, into subnetworks based on the global geometry of the network. Conceptual and quantitative ways of describing the hierarchical ordering are discussed. There are a few core-clusters centered around the most highly connected substances enclosed by other substances in outer shells, and a few other well-defined networks. This was applied to 43 organisms from the WIT database.[11] There is a strong argument for looking at the whole hierarchy tree rather than the subnetwork configuration at a specific level. For the most metabolic networks, the dominating structure at most levels of organization is the largest connected component, which means that it might be deceptive to generalize properties of subnetworks to the whole network. For the whole-cellular network, non-metabolic subnetworks, such as those representing information pathways, signal transduction and the like, are often branched from the metabolic circuitry close to the root of the hierarchy tree. Still, the largest component is dominant over a large portion of the tree's levels. Such methods should be compared to that of Seo et al.[7]

Metabolic cells are among the most modular in the cell. Accessibility of numerous complete genes as phylogenetic diverse representatives of all three kingdoms (Archaea, Bacteria, Eucarya) can help analyzing completely sequenced genomes.[12] This can help relating pathways to each other through explicit sequence information and can be particularly useful for relating electron transport in metabolic networks for the different species. Comprehension of similarities and differences between organisms and also different evolutionary rates between substrates and enzymes will follow.

**Critical Regions of Networks**

*NK* systems describe the dynamics of *N* elements upon *K* elements in the previous time step. This may result in a phase transition between a chaotic and an ordered phase. Kauffman[6] has suggested that gene networks of living organisms operate at the edge of chaos. Critical networks exhibit homeostatic stability. This implies that the connectivity in real genetic networks is not low; on the contrary, it is very high. Nonetheless, cells do not seem to operate in the chaotic phase. There are 3 phases in the dynamics of the system: a frozen phase, a marginal phase, and a chaotic phase. Evolution looks for the most stable attractors. Networks in the critical phase have the stability required to constitute evolvable systems. Their networks can recover from most mutations. Self-organized criticality (also called the edge of chaos) is found to be a paradigm in the study of models of biological processes, e.g., that of Kauffman (cellular automaton) or that of evolution according to Bak and Sneppen.[13]

Weighted network systems[14] are favored in biological systems. Spin glass models were used for inorganic molecules but biochemical molecules possess more complexity which can adapt spin glass models to reflect a higher level of complexity. Connection weights among proteins inside the cell can be incorporated in DNA using the language residing in the genetic code. Biological learning through local fitness competition modifies connection weights in response to experience. In such post genetic evolution, connection weights can be changed and implemented in the network without altering the genetic molecules permanently. This can be imprinted among hidden networks.

There is a condition for the value of the connectivity $k_c$ separating an adaptive evolutionary regime for $k>k_c$ and the "wandering" regime for $k<k_c$. $k_c$ grows with the average connectivity and "quasispecies" form at such sites that have sufficiently larger numbers of viable neighbors than average. This is purely a selective advantage: The species has a better chance to survive because it is less vulnerable to random alterations of the genetic code.[14]

We consider a lattice made up of the point and the bond as basic figures. They are denoted as ●, ○-○, ○-● and ●-●, where their symbols are *poa, p1a, pob, p1b* and *p2b*, respectively, and ○ indicates an unoccupied site, and ● indicates a site occupied by a gene, a protein or a biological molecule. The symbol between two sites indicates a bond if the two sites are both occupied. Now the ratio of occupied sites to unoccupied sites will be symbolized by *a1=p1a/poa*, *b1=p1b/pob* and *b2=p2b/pob*. It is then found that *a1=x*, *b1=x'*, and *b2=(x')2y*. The paper by Bumble and Honig[15] then adapts a systemology by introducing normalization factors, consistency relations and equilibrium relations among these geometric basic figures of the lattice to identify *x, x'*, and *y*. Introducing symbol *θ* for *p1a/(poa+p1a)* and *p1b+p2b/(poa+p1a+p2b)* for the first two approximations to the lattice we obtain the Langmuir and the Fowler-Guggenhein isotherms[16], respectively, as the first and second approximations applied to the adsorption of a gas adsorbed on a lattice. Now we can simulate this model to for of biological networks if we utilize a simple model of biomolecules as points, spheres or globules on the vertices and the bonds or edges between these vertices to be the interactions between these biomolecules or genes or proteins.

The final equations for the second approximation (using the bond as the basic figure) express the relation among chemical potential *(u)*, occupation fraction *(θ)*, coordination number *(Z)* and relative pressure *(P/P0)*. When both the coordination number and the Boltzmann factor *(c=exp(-w/kT))* are varied, the curve for the relative pressure (or chemical potential) versus the occupation number can become constant over considerable values for the occupation factor. This denotes the critical region of the graph. *Figure 1* shows the isotherm for *P/P0* when *Z=12* and *c=1.3*. This is before the critical region is found. Notice the steady rise of the curve as theta increases. (In all graphs or figures that follow, when the abscissa is denoted as theta, it denotes the occupation fraction, which is the same as the variable $\theta$.)

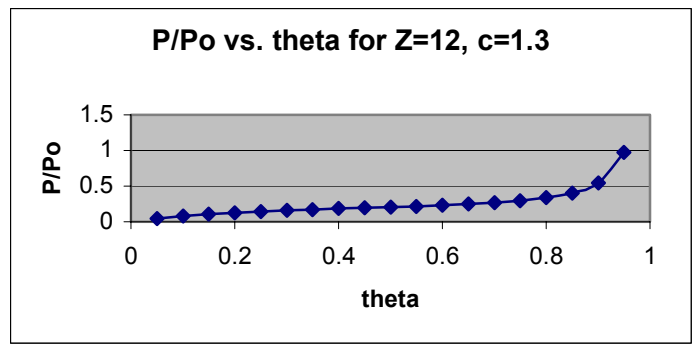

**Figure 1**

*Figure 2* shows the critical isotherm for *Z=12*, when the critical isotherm is reached at *c=1.45* and the curve becomes flat in the middle region.

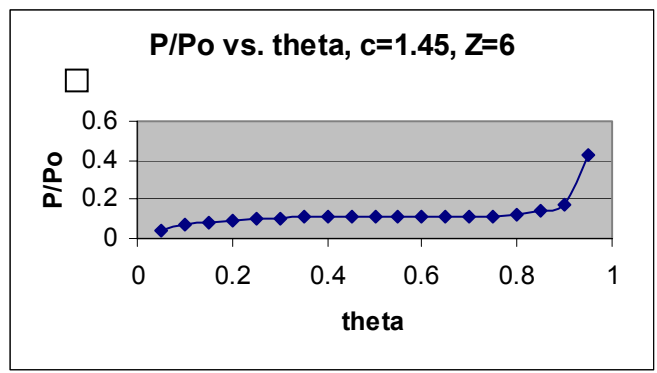

**Figure 2**

Now the curves shown use the bond as the basic figure of the network. This is the second approximation. The first approximation uses the point as the basic figure. The third approximation uses the triangle (which includes interactions between the nearest neighbors themselves) and the fourth approximation uses the rhombus as the basic figure (which includes next-nearest neighbor interactions). *Figure 3* compares the second approximation (triangle) with the second approximation (bond), and *Figure 4* compares the fourth approximation (rhombus) with the second approximation (bond). Notice that both the critical isotherm for the triangle and rhombus approximations require higher values for the interaction energy to reach the critical condition than the second approximation. In fact, the rhombus approximation occurs for the network with $Z=6$ at $c= 3.0$. The exact value due to Onsager is *3.001*.[17] The reason for resorting to the second approximation here is that the algebra for the higher approximations becomes difficult and our study here is limited to and functions as a toy model to see how the BSDK theories can be applied to cases such as the p53 connections in the mammalian gene regulatory system which has over two dozen connections.

The equations for the bond approximation to derive *Figures 1* and *2* are presented below. Calculations were repeated for various values of $Z$, $c$, and theta $(\theta)$. The results are summarized in the table below, $P/P_0=(q\exp(u/RT))=((\theta/(1-\theta))(\beta-1+2\theta)/(2c\theta))^Z$, in which $\beta=(1+4\theta(1-\theta)c(c-1))^{(1/2)}$, and $c=\exp(-w/kT)$; $R$ the gas constant; $k$ is the Boltzmann constant; and $T$ the temperature in degrees Kelvin.

| c | Z | $c=3.62Z^{-0.354}$ | Mid point $u$ | Chemical Potential - Energy | Zw |
|---|---|---|---|---|---|
| 9.1 | 3 | 2.3 | 0.7 | 10350 | 4081 |
| 4.1 | 4 | 2.1 | 0.65 | 1831 | 3477 |
| 2.8 | 5 | 1.9 | 0.5 | 204 | 3171 |
| 2.3 | 6 | 1.8 | 0.4 | -368 | 3078 |
| 1.8 | 8 | 1.6 | 0.28 | -798 | 2897 |
| 1.45 | 12 | 1.4 | 0.235 | -1028 | 2747 |
| 1.27 | 18 | 1.2 | 0.215 | -1119 | 2650 |
| 1.19 | 24 | 1.1 | 0.21 | -1153 | 2572 |
| 1.15 | 30 | 1.0 | 0.2 | -1182 | 2581 |
| 1.12 | 36 | 1.0 | 0.195 | -1179 | 2513 |

The results presented in the table are also illustrated graphically in *Figures 3* and *4*.

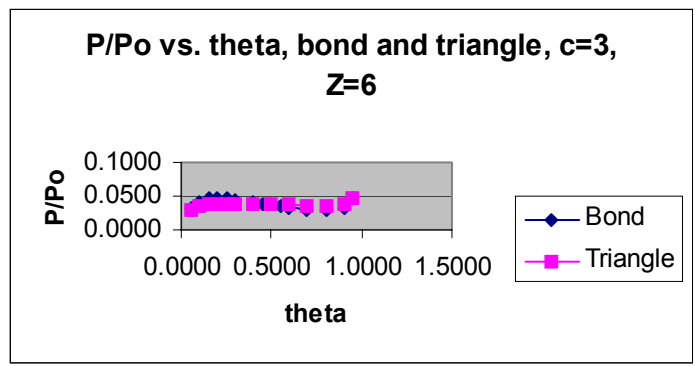

**Figure 3**

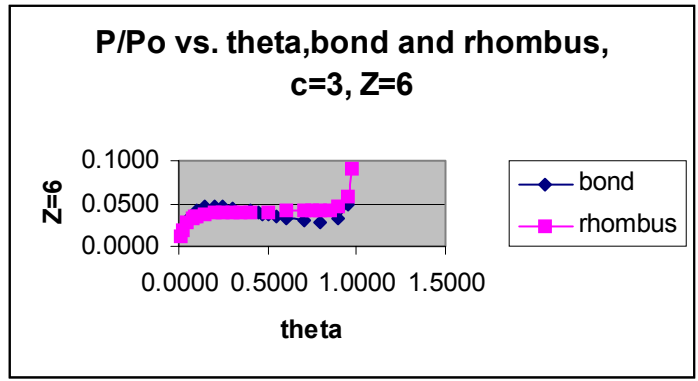

**Figure 4**

A regression upon particular numbers in the table above for $c=ac^b$ yields $c=3.62Z^{-0.354}$; see *Figure 5*. Note that the resultant correlation is practically linear for higher values of $Z$.

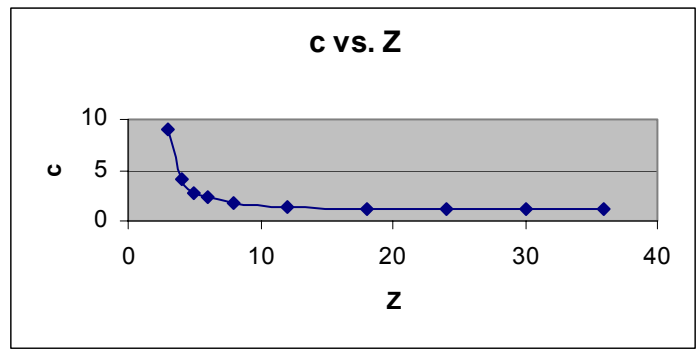

**Figure 5**

A graph of $Zw$ vs. $Z$ is plotted in *Figure 6*. Notice how the total energy of the bond with all its nearest neighbors dips and remains constant when $Z$ increases.

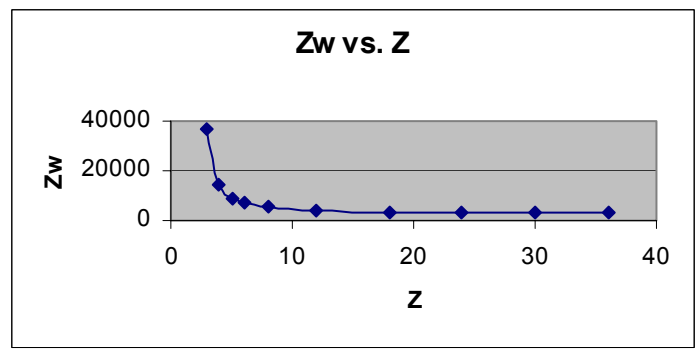

**Figure 6**

Also notice that the total interaction energy at a vertex is about 2 kilocalories, which is reasonable and that the chemical potential decreases with the increasing coordination number, thus indicating that it becomes a more stable hub. It is also noted that when the numerator of the exponent in the grand canonical ensemble *(u-E)* is graphed versus the coordination number *Z*, we obtain *Figure 7*.

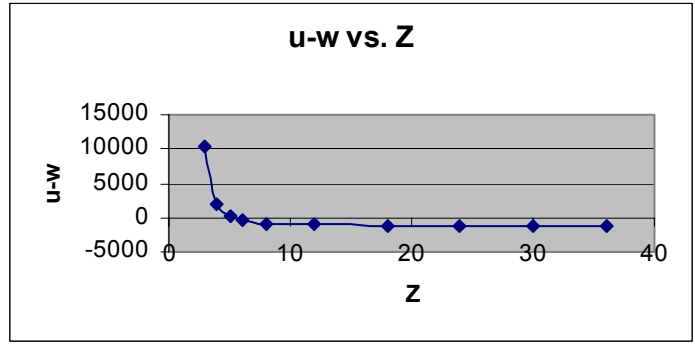

**Figure 7**

Notice in *Figure 7* that when the coordination number in the graph becomes equal to about the value of about *10* and higher, the value of the chemical potential minus the energy becomes minimum and practically stays at this value from the coordination number of up10 to the value of *36*. This indicates something about the stability of the network and also that critical regions will occur when the network may be three-dimensional. In liquids, instrumental measurements lead to a coordination number of about *10*.

In the article by Lukose and Adamic[18] it is shown that in the growth of random networks under a constraint, the diameter, defined as the average shortest path length between nodes, remains approximately constant. Since the volume of a chromosome and a cell are small, the network would have to be highly compact as it would have to fit inside the chromosome and cell.

Many approximations are involved here. One is the neglect of *q*, which is an internal partition function. Another is the use of the approximation of a bond lattice, etc. Nevertheless, the results agree with both the BSDK network ideas and also experimental values for genetic and protein networks reasonably.

The foregoing derivation of the equations for the physical properties of a lattice can also be seen as issuing from pure mathematics. The Tutte polynomial in graph theory leads to the rank generating polynomial *S(G;x,y)* of a graph *G=(V,E)*:

$$S(G;x,y) = \sum_{F > E(G)} x^{r(E)-r(F)} y^{n(F)} = \sum_{F > E(G)} x^{k(F)-k(E)} y^{n(F)}$$

Here, *S(G)=S(G;x,y)* is a polynomial in *x* and *y* and the polynomial is a function of *G*. This polynomial is useful in statistical mechanics when studying random disordered systems, especially in the neighborhood of their phase transitions. Although it is easy to give a measure proportional to the probability measure we wish to study, it is not easy to normalize it so that it becomes a probability measure. The total measure of the space (and so our normalizing factor) is the partition function. In several important cases the partition functions are simple variants of the Tutte polynomial. Additional details on these matters can be found in the book by Bollobas.[19]

On the basis of the definition of the clustering coefficient in a network, together with a regression equation for *$c=3.62k^{-0.354}$*, it is found that the density of interactions between the nearest neighbors of the centrally bonded vertex that is at its critical point can be very high indeed. If the density of interactions (bonds, sides or arcs) is very high and the density of vertices themselves is also very high, this constitutes a field. A field, similar to an electromagnetic field, can influence other fields that arise from very dense networks. This would serve to influence and integrate other biological networks. Such networks can be genetic networks, protein networks, metabolic networks, etc. This theory answers many anomalies in present day genetics. It serves to explain how the erroneous one gene ⇨ one function that came into being in the past can now be superceded. The fact that there are fewer genes and more proteins than originally thought can also be explained. The effects of these very dense fields that are very close to each other (in the cells) and acting upon each other explains the "gestalt" of an organism's functions and the way thing can fit together.

**Networks**

There has been a great interest in networks, particularly their degree, small world effect, and power law degree distribution. Growing networks exist with local rules: preferential attachment, clustering hierarchy, degree correlation and clustering hierarchy. This explosion of interest has been possible thanks to the increase of available actual network maps offering the graph representation for a wide variety of systems with sizes ranging from hundreds to billions of nodes. The degree (edges incident to a vertex) the minimum path distance between pairs of nodes and the clustering coefficient (the fraction of edges among the neighbors of a node) have attracted the attention of the physics community.

Real networks exhibit clustering or network transitivity but Erdos and Renyi's model[20] does not. The clustering coefficient, *"C"*, is the average probability that 2 neighbors of a given vertex are also neighbors of each other. The diameter *d* of a graph is the maximum distance between any 2 connected vertices in the graph. Generating function methods can generate random graphs, but when clustering is introduced, solutions become harder.

The evolution of a network can be thought of as an idealized version of a chemical process in which molecules are networks of bonds. In the evolution of networks some piece or pieces of the network could be replaced by others according to some fixed rule. Each node can be replaced at each step by a certain fixed cluster of nodes. Network systems can build up their own pattern of connections in space and time.

**Self-Organization**

There can be higher-order organizing principles that are intrinsic to nature. Self-organization is largely responsible for sculpting the biosphere. We find patterns, arising not by accident, but written into the laws of nature.

The genome is dynamic. It does not merely reflect, indirectly, the pressures of the environment; it is itself a generator of pattern and change; order from randomness can emerge. Many systems tend to organize themselves so that even with random initial conditions they end up producing behavior that has many features that are not at all random. Self-organization plays the primary role in the hierarchy of complexity. Humanity finds its place in the universe as an attractor in the space of possibilities. Given elements (such as pieces of molecules) can fit together only when certain specified constraints are satisfied. Moreover, combinatorics may cause autocatalytic systems of molecules to arise.

**Emergence**

Most interesting properties of an organism emerge from the interactions among its genes, proteins and metabolites. This implies that in order to understand the functioning of an organism, the networks of interaction involved in gene regulation, metabolism, signal transduction and other cellular and intercellular processes need be elucidated. The interactions arise from the fact that genes code, by activating or inhibiting DNA transcription. As most genetic regulatory systems of interest involve many genes, connected through interlocking positive and negative feedback loops, an intuitive understanding of their dynamics is hard to explain. In protein folding, cells assemble units called amino acids into 3 dimensional shapes that dictate the function of the resulting protein.

It is argued that the problems of emergence and the architecture of complexity can be solved by analyzing the self-organizing evolution of complex systems. Synthesizing a chemical compound is much like constructing a network by applying a specified sequence of transformations. And just like for pathway systems, it is in principle undecidable whether a given set of possible transformations can ever be combined to yield a particular chemical. The antibodies of the immune system are much like short random proteins-whose range of shapes must be sufficient to match any antigen. Some growth, particularly at a microscopic level, seems to be based on objects with particular shapes or affinities sticking together only in specific ways based on network constrained systems. The evolution of a network can be thought of as an idealized version of a chemical process in which molecules are networks of bonds. In the evolution of networks some piece or pieces of the network could be replaced by others according to some fixed rule. Each node can be replaced at each step by a certain fixed cluster of nodes. Network systems can build up their own patterns of connections in space; thus, pathway systems can also build up their own pattern of connections in time. Combinators were originally intended as an idealized way to represent structures of functions defined in logic. Many systems spontaneously tend to organize themselves, so that even with random initial conditions they end up producing behavior that has many features that are not at all random. Any rational function is the generating function for some additive cellular automaton.

**Applications**

Given elements (such as pieces of molecules) that fit together only when certain specified constraints are satisfied, it is fairly straight forward to force cellular automaton patterns to be generated (such as in the self–assembly of spherical viruses). In the existing sciences much of the emphasis over the past century or so has been on breaking systems down to find their underlying parts; then trying to analyze those parts in as much detail as possible. And particularly in physics, this approach has been sufficiently successful so that the basic components of every day systems are completely known, but just how these components act together to produce even some more of the obvious features of the overall behavior has in the past remained an almost complete mystery. There are currently about 10 million compounds listed in standard chemical databases. Of these most were identified as extracts from biological or other natural systems. In trying to discover compounds that might be useful say as drugs the traditional approach was to search large libraries of compounds, then to study variations on those that seemed promising. But in the 1980s it began to be popular to try so-called rational design in which molecules could be created that could at least to some extent specifically be computed to have relevant shapes and chemical functions. Then in the 1990s so-called combinatorial methods became popular[21], somewhat in imitation of the immune system and large numbers of possible compounds were created by successively adding at random several different possible amino acids or other units. Although it will presumably change in the future, it remained true in 2002 that half of all drugs in use are derived from just 32 families of compounds.

The p53 tumor suppressor gene may be involved in the development of about 50% of all human cancers, including breast, liver, brain, lung, colorectal, bladder and blood bourne cancers and affects approximately 125,000 patients annually in developed countries worldwide[21]. In about half of these tumors, p53 is inactivated directly as a result of mutations in the p53 gene. In many others, it is inactivated indirectly through binding to viral proteins, or as a result of alterations in genes whose products interact with p53 or transmit information to or from p53. Signaling pathways involving p53, like cellular pathways in general cannot be understood by looking at isolated components. Instead, it is essential to consider tangled networks into which these signaling components are integrated. The p53 gene is located in a gene regulatory network and is connected to over two dozen other genes. In this setting it is a hub and when activated is responsible for head and neck mutations in more that 45% to 70% of cases. Its critical function of cell cycle arrest and/or apoptosis ceases in such cases. An E1B-55kD gene-deleted adenovirus (ONYX-015) was developed for tumors lacking p53 function. The E1B-55kD gene product is responsible for p53 binding and inactivation. The E1B-55kD deletion mutant was unable to inactivate p53 in normal cells. Cancer cells lacking functional p53 (due to gene mutation) were sensitive to viral replication and cytopathic effects. Safety and tumor selective activity were indicated with intratumor injection of ONYX-15 in advanced head and neck cancer. Hence, ONYX-015 adenovirus (plus systemic cisplatin and 5-fluorouracil) provided antitumor activity and local tumor control in patients with recurrent squamous cell carcinoma of the head and neck. This is one example of a network approach to disease and treatment.

**Conclusion**

It is suggested that studies of pathways of biochemical process network syntheses of gene, metabolic and protein systems be extended to disease causes and their treatment. Furthermore, it is recommended that the examination of the order-chaos region of gene, and metabolic and protein networks be more closely explored for knowledge of the fundamental causes of diseases and their treatment.

Acknowledgement: Grateful thanks to Laszlo Papp for his proficiency and help with computer technology.

## References

1. Johnson, G., Fire in the Mind, Alfred A. Knopf, N. Y., 1995.

2. Reka Albert, Barabasi, Albert-Laszlo, Statistical Mechanics of Complex Networks, arXiv: cond-mat/0106096 v1, June 6, 2001.

3. Barabasi, Albert-Laszlo, Linked, Perseus Publishing, Cambridge, MA, 2002.

4. Steven H. Strognatz, Exploring Complex Networks, Insight Review Articles, Nature, **Vol. 410**, (pages??) March 8, 2001.

5. Dorogovtsev, S. N., Mendes, J. F. F., Evolution of Networks, arXiv: cond-mat/0106144.

6. Kauffman, S. A., Investigations, Oxford University Press, Oxford, 2000.

7. Seo, H., D-Y. Lee, S. Park, L.T. Fan, S. Shafie, B. Bertok, and F. Friedler, Graph-Theoretical Identification of Pathways for Biochemical Reactions, Biotechnology Letters, **23**, 1551-1557 (2001).

8. Fan, L.T., B. Bertok, and F. Friedler, A Graph-Theoretical Method to Identify Candidate Mechanisms for Deriving the Rate Law of a Catalytic Reaction, Computers and Chemistry, **26**, 265-292 (2002).

9. Kohn, Kurt W., Molecular Interaction Map of the Mammalian Cell Cycle Control and DNA Repair System, Molecular Biology of the Cell, **10**, 2703-2734, 1999.

10. Wolfram, S., A New Kind of Science, Wolfram Media, Inc., Champaign, 2002.

11. Holme, P., Huss, M., Jeong, H., Subnetwork Hierarchies of Biochemical Pathways, Bioinformatics, Vol.19, No. 4, pp: 532-538(7), Mar. 2003.

12. Forst, C. V., Schulten, K., Evolution of Metabolisms: A New Method for the Composition of Metabolic Pathways Using Genomics Information, J. Comput. Biol., **64**, 101-118, 1998.

13. Bak, P., Sneppen, K., Phys. Rev. Lett. **71**, 4083 (1993).

14. Vertosick, Jr., F. T., The Genius Within, Discovering the Intelligence of Every Living Thing, Harcourt, Inc., N. Y., 2002.

15. Stan Bumble, Honig, J. M., Application of Order-Disorder Theory in Physical Adsorption, I. Fundamental Equations, J. Chem. Phys., **33**, 424, (1960).

16. Fowler, R., Guggenheim, E. A., Statistical Thermodynamics, Cambridge, London, 1960.

17. Domb, C., The Critical Point, Taylor & Francis, London, 1996.

18. Lukose, R. M., Adamic, L. A., Random Networks Growing Under a Diameter Constraint, arXiv:cond-mat/0301034 v1 4 Jan 2003.

19. Bollobas, B., Modern Graph Theory, Springer-Verlag, New York, 1998.

20. P. Erdos and A. Renyi. On the Evolution of Random Graphs. Publication of the Mathematical institute of the Hungarian Academy of Sciences **5**, 17-61(1960).

21. Bumble, S., Papp, L., Fan, L. T., Friedler, F., Characteristics of Molecular-biological Systems and Process-network Synthesis, arXiv.org, physics/0203024.

22. Vogelstein, B., Lane, D., Levine, A. J., Surfing the p53 network, Nature, **408**, 307-310 (2000).